\magnification=\magstep1
\baselineskip=14pt
\overfullrule=0 pt
\font\titulobold=cmbx12 scaled\magstep1

\hfill MIT-CTP\#2706

\hfill IST-GDNL\#9/97

\hfill December~1997

\hfill {\tt arch-ive/9801055}
\vskip 1 true cm
 
\titulobold

\titulobold
\centerline{The Dynamics of Knotted Strings Attached to D-Branes} 

\tenrm
\bigskip
\centerline{\bf Ricardo Schiappa$^{a}$\footnote\dag{\rm ricardos@ctp.mit.edu}\ \ {\tenrm and}\ \ Rui Dil\~ao$^b$\footnote\ddag{\rm rui@sd.ist.utl.pt}}
\bigskip
\centerline{\it a-  Center for Theoretical Physics and Department of Physics}
\centerline{\it Massachusetts Institute of Technology, 77 Massachusetts Ave.}
\centerline{\it Cambridge, MA 02139, U.S.A.} 
\medskip
\centerline{\it b-  Grupo de Din\^amica N\~ao-Linear, Departamento de F\'\i sica}
\centerline{\it Instituto Superior T\'ecnico, Av. Rovisco Pais}
\centerline{\it 1096 Lisboa Codex, Portugal}

\bigskip
\bigskip
\bigskip
\bigskip

\centerline{\bf Abstract}

We extend the general solution to the Cauchy problem for the relativistic 
closed string (Phys. Lett. {\bf B404} (1997) 57-65) to the case of open strings attached to D$p$-branes, including the cases where the initial data has a knotlike 
topology.  We use this extended solution 
to derive intrinsic dynamical properties of open and closed relativistic strings
attached to D$p$-branes. 
We also study the singularity structure and the oscillating periods
of this extended solution.

\vfill


\eject

String  theory is 
 an important field of research
due to the rich phenomena and physical concepts associated to nonlinear 
field theories.
Pioneering work in general nonlinear
field theory goes back to  Born and Infeld where the two-dimensional
scalar equation of  nonlinear electrodynamics is the equation of motion for a
string in Minkowsky space [1-3]. 
More recently, it has been recognized that spinning
massless strings of finite lenght  in Minkowsky space behave like particles with
nonzero rest mass and intrinsic spin [4-7].

In the general context of nonlinear field theories, it has been analytically shown that knotlike string configurations appear as stable solutions of the evolution 
equations [8]. Recently, Faddeev and Niemi [9-10] have shown, with methods of 
high performance computing, that knotlike configurations appear as stable solitons 
in certain relativistic quantum field theories.

Several problems related with the dynamical evolution of  strings
and in particular knotted strings remain unsolved.  In the context of the theory
of galaxy formation, strings formed at a phase transition verly early in the history of the universe might provide the density perturbations to start the condensation
process. Different estimates  for the temperature transitions exist depending
on the dynamical configuration of strings: strings with a random configuration [11]
or strings with stable closed loops [12].

All these questions have an answer if  explicit solutions for nonlinear
equations of motion supporting arbitrary string configurations and 
topologies at some initial time
exist.
In fact,  the general Cauchy
problem for the free motion of a closed string, knotted or not,  in the Minkowski space $M^{3+1}$ has been solved [8]. It has been shown that  (1) Initially static closed strings 
always oscillate 
with the period $T=\ell /2c$, where $\ell $ is the string length;  
(2) Oscillating knotted strings show singular behavior leading
to the simple link for times $t=(2 n+1)\ell /4c$, with $n=0,1,\ldots$;
(3) Knotted strings can stretch infinitely; 
(4) Infinite strings have solitonic like behaviour in the sense that localized 
string perturbations propagate without loosing coherence [13]. 
In the case of  collision
of two opposite velocity packets the coherence is lost, but after collision both packets aquire their initial
shape and  propagate freely with the same  velocity. 

When dealing with open strings, boundary conditions on the coordinates must be considered, with the possibility of choosing these as either of Dirichlet or Neumann type. Appropriately 
choosing these boundary conditions, we can construct D-branes, which are extended objects 
with the property that strings can end on them (see [14-15], and references therein). 
One important feature of D-brane theory relies on the hypothesis that if black holes 
are described as a collection of D-branes with attached strings, 
the 
Bekenstein-Hawking entropy computed from the classical black hole solution agrees with
the black hole entropy 
given by the number of states of strings and D-branes (see [16] and 
references therein). A general proof of this result in four dimensional spacetime 
is, however, still lacking.

The aim of this paper is to describe the dynamics of strings 
attached to D$p$-branes in Minkowsky space $M^{3+1}$, with $p=0,1,2,3$,  
including knotted string configurations.

\bigskip

We consider strings in Minkowsky space $M^{3+1}$  with metric
$d\ell^2=dt^2-dx^2-dy^2-dz^2=(dx^0)^2-(dx^1)^2-(dx^2)^2-(dx^3)^2$, and units
such that $c=1$. The motion of strings in $M^{3+1}$ is described by the 
Nambu action 
$$
A=-m\int\int dp\ dq\ \sqrt{
(\partial_p {{\bf r}}\cdot \partial_q {{\bf r}})^2 -
(\partial_p {{\bf r}})^2 (\partial_q {{\bf r}})^2}:=\int dp\ dq\ {\cal L}\eqno(1)
$$
where $p$ and $q$ parameterize the surface $S$ spanned by the string in $M^{3+1}$,  $m$ is a constant of proportionality, 
${{\bf r}}=(x^0(p,q),x^1(p,q),x^2(p,q),x^3(p,q))$ is the four dimensional
position vector and   ${\cal L}$ is the Lagrangean density of the string.

As the $p$ and $q$ parameters are arbitrary, we can choose  $p$ and $q$ such that, [2] and [8],
$$
(\partial_p {{\bf r}})^2=0\, ,\quad (\partial_q {{\bf r}})^2=0\eqno(2)
$$
where $p=p(t,\sigma )$ and $q=q(t,\sigma )$. The parameter $\sigma$ describes the
string shape in the proper reference frame of the string. 
Under these conditions,  the equation 
of motion for string perturbations is
$$
\partial_p \partial_q {{\bf r}}=0\eqno(3)
$$
provided the constraint equations (2) hold.

For initial string positions and velocities described by functions $a^k(\sigma)$ and $b^k(\sigma)$ such that
$$\eqalign{
x^{k}(0,\sigma)&=a^{k}(\sigma)\cr
\partial_tx^{k}(t,\sigma)|_{t=0}&=b^{k}(\sigma)\, , \quad k=1,2,3\cr}\eqno(4)
$$
the general solution of the Cauchy problem, including the cases
of knotted topologies, is,  [8],
$$\eqalign{
t(p,q)&=  {1\over 2}\int_p^q  \Pi (s) ds\cr
x^k(p,q)&={1\over 2} (a^k(p)+a^k(q))+
{1\over 2}\int_p^q  b^k(s)\Pi (s) ds
\, ,\   k=1,2,3\cr}\eqno(5)
$$
where
$$
\Pi (s)=\left({ 
{\sum_{i=1}^3 (\partial_{s} a^i(s))^2} \over
1-\sum_{i=1}^3(b^i (s))^2
}\right)^{1/2}\eqno(6)
$$
provided 
$$\eqalignno{
&0\le \sum_{i=1}^3(b^i (\sigma))^2<1&(7a)\cr 
&\sum_{i=1}^3 b^i (\sigma) {\partial_{\sigma}a^i(\sigma)}=
(\partial_t {{\bf r}}\cdot \partial_{\sigma} {{\bf r}})|_{t=0}=0&(7b)\cr 
}
$$
and $p=q=\sigma$, for $t=0$. 
In the string proper reference frame $(t,x^1,x^2,x^3)$,
 the Minkowsky distance between string points is negative. Therefore,
the string lies on a space-like three-dimensional surface in $M^{3+1}$. 

One of the consequences of the solution of the Cauchy problem for the relativistic
string is related with the impossibility of having infinite or closed 
rotating  strings, {\it defined for all times}. We exemplify this effect with 
two examples. Suppose that we have initial Cauchy data, $a^1(\sigma )=\cos (\sigma)$,
$a^2(\sigma )=\sin (\sigma)$, $a^3(\sigma )=0$, $b^1(\sigma )=-\omega \sin (\sigma)$,
$b^2(\sigma )=\omega \cos (\sigma)$ and $b^3(\sigma )=0$, representing a 
rotating closed loop with angular velocity $\omega $,
at $t=0$. For the Cauchy problem
to be well defined we must verify (7). But by (7b),  it follows that
$\omega =0$, and we cannot have simple closed 
rotating relativistic strings.

In the case of an infinite string we take as initial data
$a^1(\sigma )=\sigma$,
$a^2(\sigma )=0$, $a^3(\sigma )=0$, $b^1(\sigma )=0$,
$b^2(\sigma )=  f(\sigma)$ and $b^3(\sigma )=0$, where $f(\sigma)$ is a  strictly
increasing even function. By (7a), we obtain the condition $|f(\sigma)|<1$,
which is not compatible with the initial requirement for the function 
$f(\sigma)$. For finite open strings, $f(\sigma)$ must be extended to 
$\sigma \in {\bf R}$ (see below), and so we still cannot have rotation.

We note that, according to Scherk [5], we can relate the classical spin of the 
string, $J$, to the mass squared of the string, $M^2$, through the inequality 
$J\le\alpha'M^2$, where $\alpha'$ is a constant related to $m$ in (1). Moreover, the 
equality is reached only for a rigid rotating string. Thus, our previous discussion 
shows that the equality is not compatible with the solution of the Cauchy 
problem.

Choosing the new string world sheet parameterization,
$$
\nu ={p+q\over 2}\, , \quad \tau={q-p\over 2}\eqno(8)
$$
string solutions  are now
$$\eqalign{
t(\tau,\nu)&=
{1\over 2}\int_{\nu -\tau}^{\nu +\tau} \Pi (s) ds\cr
x^k(\tau,\nu)&={1\over 2} (a^k(\nu -\tau)+a^k(\nu +\tau))+
{1\over 2}\int_{\nu -\tau}^{\nu +\tau}  b^k(s)\Pi (s) ds
\, ,\   k=1,2,3\cr }\eqno(9)
$$
In the new parameters $(\tau, \nu)$,
the equation of motion of the string becomes
$$
(\partial_{\tau\tau}- \partial_{\nu\nu} ){{\bf r}}=0\eqno(10)
$$
with constraints
$$
(\partial_{\nu} {{\bf r}})^2+(\partial_{\tau} {{\bf r}})^2=0\, ,\quad 
\partial_{\nu} {{\bf r}}\cdot \partial_{\tau} {{\bf r}}=0\eqno(11)
$$
 For $t=0$ and $p=q=\sigma$, $\tau=0$ and
$p=q=\sigma=\nu$.

We now consider the case where strings have finite lengths and the end points 
are connected to a $p$-dimensional hypersurface, a D$p$-brane, with $p=0,1,2,3$.
To fix ideas, we first consider the case where strings 
have fixed extreme points --- the D0-brane. In this case,
we consider that string extreme points occur for
$\nu=0$ and $\nu=\pi$. The motion of the attached string can then be discribed
by a set of initial position and velocities, $\bar a^k(\nu)$ and $\bar b^k(\nu)$
as in (4), with $\nu \in [0,\pi ]$, together with the spatial Dirichlet boundary 
conditions:
$$
x^k(\tau,\nu=0)=X^k_1(\cdot)\, ,\quad  x^k(\tau,\nu=\pi)=X^k_{2}(\cdot)\, ,\   k=1,2,3\eqno(12)
$$ 
where $X^k_1(\cdot)$ and $X^k_{2}(\cdot)$ are arbitrary functions associated to the D0-branes. For example, for a static D$0$-brane, the functions
$X^k_{i}(\cdot)$ are constants. For a dynamic D$0$-brane, we can have $X^k_{i}=X^k_{i}(\xi)$, where $\xi$ is a new parameter describing the
dynamics of the attachement points. In general, we can assume that $\xi=\xi(\tau)$.

For a D$3$-brane, the string has moving end points and we must choose 
spatial Neumann boundary conditions
$$
\partial_{\nu}x^k(\tau,\nu)|_{\nu=0}=0\, ,\quad  \partial_{\nu}x^k(\tau,\nu)|_{\nu=\pi}=0\, ,\   k=1,2,3\eqno(13)
$$ 

In Dirichlet, Neumann and mixed cases, we have an additional Neumann boundary
condition on the temporal coordinate, [14-15],
$\partial_{\nu}t(\tau,\nu)|_{\nu=0,\pi}={1\over 2}(\Pi(\nu+\tau)-\Pi(\nu-\tau))|_{\nu=0,\pi}=0$. As we will see below, this is always true as $\Pi(s)$
is an even function around the extreme points of the string. Therefore,
a generic D$p$-brane has $p+1$ Neumann boundary conditions 
 and $(3-p)$ 
Dirichlet boundary conditions, in the $(\tau, \nu)$ parameterization.

Let us now extend the general solution (9) of the Cauchy problem for the
D$0$-brane case.
Suppose  that  (9)  holds in the 
finite lenght string case. Therefore, by (12), for $\nu =0$ and $\nu= \pi$,
$$\eqalign{
x^k(\tau,0)&={1\over 2} (a^k(-\tau)+a^k(\tau))+
{1\over 2}\int_{-\tau}^{\tau}  b^k(s)\Pi (s) ds=X^k_1(\xi)\cr
x^k(\tau,\pi)&={1\over 2} (a^k(\pi-\tau)+a^k(\pi+\tau))+
{1\over 2}\int_{\pi-\tau}^{\pi+\tau}  b^k(s)\Pi (s) ds=X^k_2(\xi)\, ,\  
 k=1,2,3\cr}\eqno(14)
$$
Suppose now that, $b^k(s)\Pi(s)$ is an odd function around the two extreme points 
of the string, $s=0$ and $s=\pi$. Then, relations (14) simplify to
$$\eqalign{
a^k(-\tau)&=-a^k(\tau)+2X^k_1(\xi)\cr
a^k(\tau+\pi)&=-a^k(\pi-\tau)+2X^k_2(\xi)\, ,\   k=1,2,3\cr}\eqno(15)
$$
for every $\tau$. As (15) must hold for $\tau \in {\bf R}$, the functions 
$a^k(\nu)$  can be constructed if we know the values of  $a^k(\nu)$ in
the interval $[0,\pi]$. Therefore,
given the initial string position $\bar a^k(\nu)$,
with $\nu \in [0,\pi]$, $\bar a^k(0)=X^k_1$ and $\bar a^k(\pi)=X^k_2$, it is always possible
to extend $\bar a^k(\nu)$ to $a^k(\nu)$ as  odd functions around the points $X^k_1$ and $X^k_2$.  This extension is easily obtained with $a^k(\nu)=\bar a^k(\nu)$ for
$\nu \in [0,\pi]$, relations (15) with $\tau \to \nu$, and
$$
a^k(\nu+n\pi)=a^k(\nu+(n-2)\pi)-2 X^k_1+2X^k_2\, ,\   k=1,2,3\eqno(16)
$$ 
derived by induction from (15).

As $a^k(\sigma)$ is an odd extension of $\bar a^k(\sigma)$, 
and if $b^k(\nu)$ is  an odd extension of $\bar b^k(\nu)$, with
$b^k(\nu=0)=b^k(\nu=\pi)=0$,
by (6),
$\Pi(s)$ is even. Therefore,  $b^k(s)\Pi(s)$ is also an odd 
function around the extreme points of the string, as we have initially assumed.

Hence, we have proved that for given initial functions $\bar a^k(\nu)$ and $\bar b^k(\nu)$, the  string attached to a static or dynamic D$0$-brane
evolves according to solutions (9), where $a^k(\nu)$ and $b^k(\nu)$
are odd extensions of $\bar a^k(\nu)$ and $\bar b^k(\nu)$ around
the string end points and $b^k(0)=b^k(\pi)=0$. These odd extensions are calculated 
through (15) and (16).

For the case of a string attached to a D$3$-brane we impose the Neumann boundary conditions (13) to the solution (9) for the motion of the relativistic string. 
Introducing (9) into (13), we obtain,
$$\eqalign{
{\dot a}^k(-\tau)+{\dot a}^k(\tau)+b^k(\tau )\Pi (\tau )-
b^k(-\tau )\Pi (-\tau )&=0\cr
{\dot a}^k(\pi-\tau)+{\dot a}^k(\pi+\tau)+b^k(\pi+\tau )\Pi (\pi+\tau )-
b^k(\pi-\tau )\Pi (\pi-\tau )&=0\, ,\  
 k=1,2,3\cr}\eqno(17)
$$
where,  ${\dot a}^k(-\tau)=\partial_s a^k(s)|_{s=-\tau}$. 
Now we consider that the initial Cauchy data is specified by functions
${\bar a}^k(\nu )$ and ${\bar b}^k(\nu )$, with $\nu \in [0,\pi]$, as in (4).
Let us suppose further  that the product functions $b^k(\nu )\Pi(\nu )$
are even around the points $\nu =0$ and $\nu =\pi$. Hence, by (17),
$$\eqalign{
{\dot a}^k(-\tau)&=-{\dot a}^k(\tau)\cr
{\dot a}^k(\tau+\pi)&=-{\dot a}^k(\pi-\tau)\cr
{\dot a}^k(\tau+n\pi)&={\dot a}^k(\tau+(n-2)\pi)\, ,\   k=1,2,3\cr
}\eqno(18)
$$
But if ${\dot a}^k$ are odd functions around $\tau =0$ and $\tau=\pi$, 
as (18) implies, we can construct $a^k(\nu)$ by the even extension of 
${\bar a}^k(\nu )$ around the points $\nu =0$ and $\nu =\pi$. Therefore by (18)
we have
$$\eqalign{
a^k(-\nu)&=a^k(\nu)\cr
a^k(\nu+\pi)&=a^k(\pi-\nu)\cr
a^k(\nu+n\pi)&=a^k(\nu+(n-2)\pi)\, ,\   k=1,2,3\cr
}\eqno(19)
$$
where $a^k(\nu)={\bar a}^k(\nu )$ for $\nu \in [0,\pi]$. Under these conditions, 
with $b^k(\nu)$ even, around  $\nu =0$ and $\nu =\pi$, the function 
$b^k(\nu )\Pi(\nu )$ is
also even around the extreme points of the string, as we have initially assumed.

Therefore,  (9) is also a solution of the D$3$-brane problem,
where $a^k(\nu)$ and $b^k(\nu)$
are even extensions of $\bar a^k(\nu)$ and $\bar b^k(\nu)$ around
the string end points. The even extension of the initial functions are calculated according to the recurrence relations (19), for both $a^k(\nu)$ and $b^k(\nu)$.

The general case for D$p$-branes, with $p=1,2$ is simply calculated through
(15), (16) and (19),  with $p+1$ Neumann  and $(3-p)$ 
Dirichlet boundary conditions.

In Figs. 1, 2, 3 and 4 we represent the time evolution of finite length strings attached to D$0$, D$2$ and D$3$-branes. The numerical implementation of the
solutions of the Cauchy problem for initially static strings shows that 
we have always periodic motion with periods $T=\ell$, Figs. 1 and 4,  or, 
$T=2\ell$,
Figs. 2 and 3, where $\ell$ is the string length at $t=0$. This differs from the free closed string case where the period is always $T=\ell /2$, [8]. 

The singularity structure of string solutions is easily analysed with
the techniques developed in [8]. In the case of finite open strings attached 
to two D$0$-branes no singularities occur, Fig. 1, as in the case of a finite 
free open string (attachement to a D$3$-brane), Fig. 4. However, 
in the cases of Figs. 2 and 3, singular solutions occur for   times
$t=(2 n+1)T/4$, with $n\ge 0$, as already observed in the free closed string case.

In Fig. 1, the initial string configuration has an expansion in Fourier
series with only one non-zero eigenmode. However, for $t>0$, as clearly seen
from the figures, the  eigenmodes  become time $t(\tau ,\nu)$ dependent
and the energy flows from eigenmode to eigenmode, with periodicity $T$.
This could have consequences for the quantization of open strings [7].
Generically,
strings attached to D$2$-branes, Fig. 3, intersect  the brane, introducing
some difficulties in the analysis of systems made of strings attached 
to D$p$-branes, namely, we would need dynamical branes and/or interaction
terms between the brane and the string.

General properties extracted from the figures indicate that for "closed`` 
strings (where the closure is performed by the brane) the period is 
$T=2\ell$ and a singularity occurs at $t=\ell /2=T/4$.
In the case of Figs. 2a) and 3, the singularity is a single point; while in
the case of Fig. 2b), the treefoil knot, the singularity is a collection of three 
points: the projection of the knot onto a planar surface. For "open`` strings
the period is $T=\ell$ and no singularities seem to occur, Figs. 1 and 4.

In conclusion, we have solved explicitly the Cauchy problem for 
strings attached to D$p$-branes, with $p=0,1,2,3$. Initially static strings attached to D$p$-branes oscillate with periods $T=\ell/ c$ or $T=2\ell/ c$,
contrasting with the period $T=\ell/ 2c$ for the free closed string. 
Singularities occur throughout string motion, with more complex patterns 
than in the free closed string.

\bigskip

{\bf Acknowledgments:}  
We would like to thank Jo\~ao Nunes for suggestions and a careful
reading of the manuscript.
One of us (RS) is partially supported by the
Praxis XXI grant BD-3372/94 (Portugal).

\vfill \eject
\centerline{\bf References}
\bigskip 

[1] --- M. Born \& L. Infeld, {\it Foundations of the New Field Theory}, 
Proc. Roy. Soc. {\bf A144} (1934) 425-451. 

[2] --- B. M. Barbashov \& N. A. Chernikov, {\it Solution and Quantization 
of a 
Nonlinear Two-Dimensional Model for a Born-Infeld Type Field},
Soviet Phys. JETP {\bf 23} (1966) 861-868. 

[3] --- B. M. Barbashov \& N. A. Chernikov, {\it Solution of the
Two Plane Wave Scattering Problem in a Nonlinear Scalar Field 
Theory of the   Born-Infeld Type}, Soviet Phys. JETP {\bf 24} 
(1967) 437-442. 

[4] --- P. Goddard, J. Goldstone, C. Rebbi \& C. B. Thorn, {\it Quantum Dynamics
of a Massless Relativistic String}, Nuc. Phys. {\bf B56} (1973) 109-135.

[5] --- J. Scherk, {\it An Introduction to the Theory of Dual Models and 
Strings}, Rev. Mod. Phys. {\bf 47} (1975) 123-164.

[6] --- B. Felsager, {\it Geometry, Particles and Fields},
Odense Univ. Press (1981). 

[7] --- M. B. Green, J. H. Schwarz \& E. Witten, {\it Superstring
Theory}, Cambridge Univ. Press (1987).

[8] --- R. Dil\~ao  \& R. Schiappa, {\it Stable Knotted Strings}, 
Phys. Lett. {\bf B404} (1997) 57-65, hep-th/9704084. 

[9] --- L. Faddeev \& A. J. Niemi, {\it Stable Knot-Like Structures in
Classical Field Theory}, Nature {\bf 387} (1997) 58-61,
hep-th/9610193. 

[10] --- L. Faddeev \& A. J. Niemi, {\it Toroidal Configurations as Stable Solitons}, 
hep-th/9705176.

[11] --- Ya. B. Zeldovich, {\it Cosmological Fluctuations Produced near a Singularity}, 
Mon. Not. R. Astron. Soc. {\bf 192} (1980) 663-667.

[12] --- A. Vilenkin, {\it Cosmological Density Fluctuations Produced by Vacuum 
Strings}, Phys. Rev. Lett. {\bf 46} (1981) 1169-1172.

[13] --- S. Brundobler \& V. Elser, {\it Colliding Waves on a Relativistic 
String}, Am. J. Phys. {\bf 60} (1992) 726-732. 

[14] --- J. Polchinski, S. Chaudhuri \& C. V. Johnson, {\it Notes on D-Branes}, 
hep-th/9602052.

[15] --- J. Polchinski, {\it TASI Lectures on D-Branes}, hep-th/9611050.

[16] --- G. P. Collins, {\it Quantum Black Holes are Tied to D-Branes 
and Strings}, Phys. Today {\bf 50} number 3 (1997) 19-22.

\vfill \eject

\centerline{\bf Figure Captions}
\bigskip

{\bf Figure 1:} Time evolution of a finite length string attached to two D$0$-branes,
and initial data $a^1(\nu )=\nu$, $a^2(\nu )=\sin (2\nu)$, $a^3(\nu )=0$
and $b^k(\nu )=0$. The motion is periodic with period $T=\ell$, where $\ell$
is the string length at $t=0$. The two D$0$-branes are represented by large
black dots with static space coordinates $(X_1^1=0,X_1^2=0,X_1^3=0)$ and $(X_2^1=1,X_2^2=0,X_2^3=0)$.
We display a tubular neighborhood  around the 
string in order to better depict the knot topology.
\bigskip

{\bf Figure 2:} Time evolution of closed strings attached to a D$0$-brane.
The motion is   periodic with period  $T=2\ell$, where $\ell$
is the string length at $t=0$. In case a) the initial string configuration is
a closed loop and the string solutions shows a singularity for $t=\ell /2$.
In case b) the initial string configuration is a treefoil knot and during
time evolution the string solution has several singularities corresponding
to the crossover of string strands, and in particular it also shows a 
singularity for $t=\ell /2$.

\bigskip

{\bf Figure 3:} Time evolution of a finite length string attached to a D$2$-brane
in the $(x,y)$ plane,
and initial data $a^1(\nu )=\cos (\nu)$, $a^2(\nu )=\cos (\nu)$, $a^3(\nu )=\sin (\nu)$
and $b^k(\nu )=0$. The motion is periodic with period $T=2\ell$. For
$t=\ell /2$, the string solution has a singularity, collapsing into the D$2$-brane.

\bigskip

{\bf Figure 4:} Time evolution of a finite length string attached to a D$3$-brane.
 The motion is periodic with period $T=\ell$.

\vfill

\bye